\documentclass[letterpaper]{article}
\usepackage[utf8]{inputenc}
\usepackage{url}
\usepackage{isea}
\usepackage[pdftex]{graphicx}
\usepackage{times}
\usepackage{helvet}
\usepackage{courier}
\usepackage[final]{changes}
\usepackage{caption}
\usepackage{subcaption}
\usepackage{mciteplus}
\usepackage{hyperref}
\usepackage{listings}
\usepackage[frozencache=true,cachedir=minted-cache]{minted}
\usepackage[numbers]{natbib}

\newcommand{\mycomment}[1]{}

\title{Lightweight Implementation of Per-packet Service Protection in eBPF/XDP}
\author{ 
	Ferenc Fejes\textsuperscript{1}, Ferenc Orosi\textsuperscript{2}, Balázs Varga\textsuperscript{1}, János Farkas\textsuperscript{1}\\
	\textsuperscript{1} Ericsson Research TrafficLab, \textsuperscript{2} Eötvös Loránd University\\
	Budapest, Hungary\\
	ferenc.fejes@ericsson.com\\
	\newline
	\newline
}

\begin{document} 
\maketitle




\begin{abstract}
Deterministic communication means reliable packet forwarding with close to zero packet loss and bounded latency.
Packet loss or delay above a threshold caused by, e.g., equipment failure or malfunction could be catastrophic for applications that require deterministic communication.
To meet loss related targets, per-packet service protection has been introduced by deterministic communications standards;
it is provided by Frame Replication and Elimination for Reliability (FRER) for Layer 2 Ethernet networks and by Packet Replication, Elimination, and Ordering Functions (PREOF) for Layer 3 IP/MPLS networks.

We have implemented FRER with two conceptually different methods: 
(1) in eBPF/XDP as a lightweight software implementation; 
and (2) in userspace.
We evaluate our XDP FRER via an experimental analysis and compare the two FRER implementations.
\end{abstract}

\maketitle

\section{Introduction}

Several applications - such as industrial automation, aerospace, automotive - require deterministic communication, where high reliability, bounded latency, low packet delay variation (jitter), and no packet loss due to congestion or network failure are essential.
This implies that the network should be carefully engineered and provisioned by the operator to avoid congestion-related packet losses or high queueing delays.
Time-sensitive applications typically communicate with periodical data transmission, therefore, their traffic characteristics are known by the operator.
However, equipment failures can happen even in the most carefully engineered networks, potentially resulting in communication outage.
In an industrial environment, the wiring may break, the radio link can be temporarily shielded by an obstacle, or the network nodes can be damaged too. 

Many legacy networks apply resiliency techniques like rerouting to cope with topology changes, e.g., failures.
However, convergence of a routing protocol takes time, which causes outage in the communication.
To further mitigate the impact of failures, protection switching techniques are used to reduce the failover time.
However, protection switching techniques are dependent on failure detection and signaling of switchover, so they cannot fully eliminate possible outage during network transition.
Per-packet service protection has been introduced to overcome the failover time problem mentioned above for former methods. 

Per-packet service protection is specified for Layer 2 networks by the IEEE 802.1 Time-Sensitive Networking (TSN) Task Group in the IEEE 802.1CB \textit{Frame Replication and Elimination for Reliability}~\cite{1cb} (FRER) standard.
Along a common architecture with TSN, the IETF Deterministic Networking Working Group defines per-packet service protection for Layer 3 networks as \textit{Packet Replication, Elimination and Ordering Functions} (PREOF) see, e.g., RFC 8655~\cite{rfc8655}.
The operation of FRER and PREOF Replication (R) and Elimination (E) functions are essentially the same.
They rely on (maximally disjoint) redundant forwarding paths, which are explicit, to avoid convergence issues.
As they are fixed paths (a.k.a. ''nailed-down paths''), they do not change even in case of a change in the network topology.
The Replication function replicates packets, amends their header with a sequence number to be able to identify copies of the same packet and transmits the replicas to redundant network paths.
The Elimination function forwards one of the replicas of a packet and discards the surplus.
The IEEE specification dives into the deep details of the operation; therefore, our paper focuses on FRER. 

The rest of the paper is structured as follows.
First, we briefly describe the per-packet protection operation, with the help of an example network and a failure scenario.
After that, we present our lightweight eBPF/XDP~\cite{xdp} implementation of FRER, referred to as \textit{XDP FRER} in the rest of the paper, and discuss its benefits and limitations.
Finally, we evaluate XDP FRER and compare it with an alternative, userspace FRER implementation (referred to as \textit{uFRER}).

\section{Per-Packet Service Protection}

The two main components of the per-packet service protection are the replication and elimination functions.
Their operation requires the identification of streams (data flows) a packet belongs to.
Streams are identified based on the packet header fields, specified in the TSN/DetNet standards.
The Layer 2 and Layer 3 standards specified different header fields for stream identification at the different layers.
A given per-packet service protection implementation has to support matching for these header fields and is configured which values are used for the identification of a given stream.
It can be configured which header fields are used for stream identification. 

At the edge of the network a unique sequence number is generated for each packet of a stream. 
The sequence number increases monotonically and warps around when it reaches its maximum value.
The sequence number is encapsulated in an appropriate header field (e.g., redundancy tag); thus, it is carried with the packet across the network.
The format of that header depends on whether Layer 2 or Layer 3 service protection is used.

The node running the elimination function receives the packets and identifies whether they belong to a protected stream.
The elimination function decides based on the sequence number of the packet about forwarding.
Only the replica received first is forwarded; every other replica packet is dropped.
A so-called history window stores which sequence numbers have been recently received, hence it is the basis for the drop of replicas.
The elimination algorithm is stateful, it stores the recently received sequence numbers of the served stream, what is the basis for the drop of replicas.

A benefit of this operation is that the decision about the received packet is immediate - no buffering is necessary.
Note: deterministic streams are uni-directional.
Replication/Elimination functions are stream specific.

\subsection{Operation example}

\begin{figure}[tp]
\centering
\includegraphics[width=0.95\columnwidth]{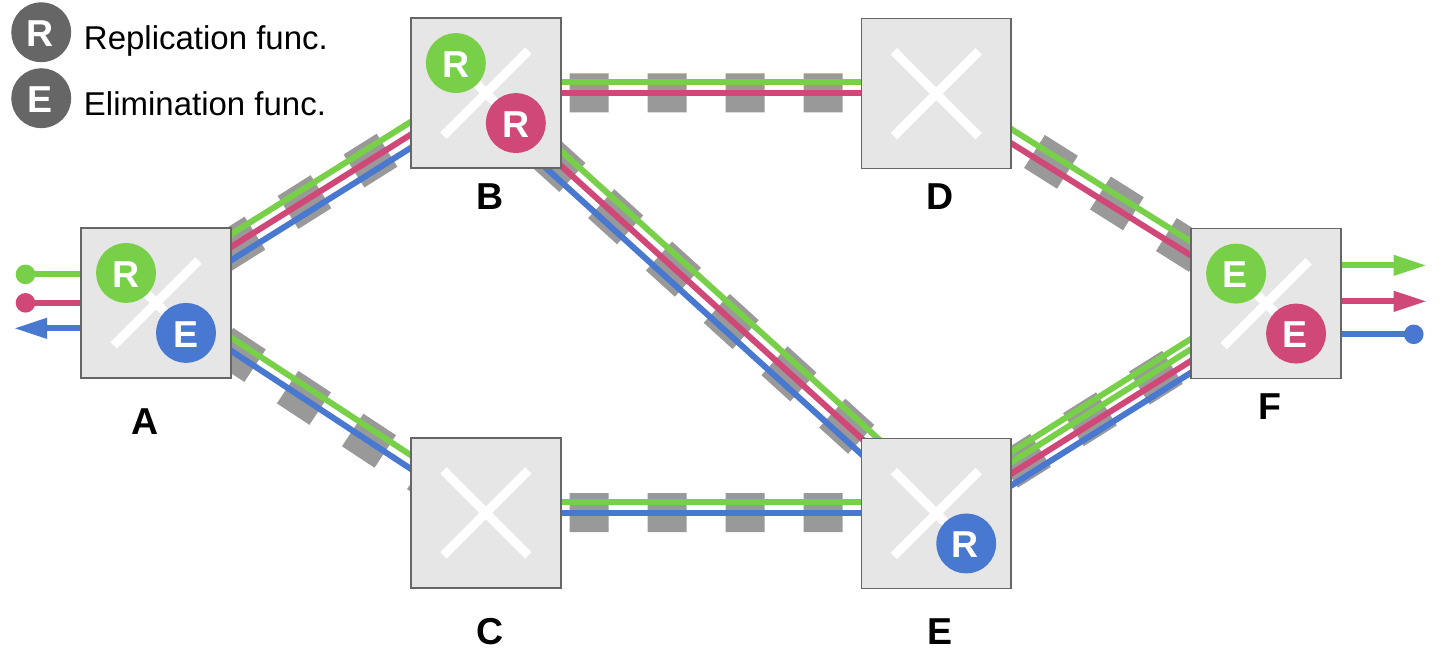}
\caption{Example network with three streams, six networking nodes and multiple replication and elimination function instances}
\label{fig:frerexample}
\end{figure}

The example network in Figure \ref{fig:frerexample}. show three streams, \textbf{red}, \textbf{green}, and \textbf{blue}, and six network nodes \textbf{A-F}.
The letter \textbf{(R)} indicates the replication and \textbf{(E)} the elimination functions for the streams (they color-coded according to their corresponding stream).
Stream \textbf{blue} is protected by the replication function on \textbf{node-E} sending the replicas on disjoint paths via \textbf{B} and \textbf{C}.

Similarly \textbf{green} is protected by the replication function on \textbf{node-A}.
Multiple replication functions can enhance the protection of a stream, e.g. \textbf{green} is further protected by its replication function on \textbf{node-B}.
\textbf{Red} is protected by its replication function on \textbf{node-B}.
The elimination functions on \textbf{node-F} can handle arbitrary replicas of the same packet.

In case the \textbf{B-E} link fails, all of the streams continue to operate without packet loss.
Furthermore, as one can see, none of the streams are affected by individual link failures of \textbf{A-C}, \textbf{C-E}, \textbf{D-F}, or \textbf{B-D}.

Depending on the network topology and the placement of the replication and elimination functions, further protection could be applied.
In our example, \textbf{red} and \textbf{blue} streams can be further protected by configuring replication functions on nodes \textbf{node-A} and \textbf{node-F} accordingly. 

In our paper, we investigate the implementation of the Layer 2 per-packet service protection functions i.e., FRER.
Implementation of Layer 3 PREOF is similar from functionality perspective, but not discussed here.

\section{Frame Replication and Elimination for Reliability}

FRER has its standard encapsulation for the sequence number, which is called Redundancy-tag (R-tag). R-tag is a six-byte field, which includes a 2-bytes sequence number.

The FRER functions can be implemented in locations where Layer 2 packet processing takes place (e.g. hardware, kernelspace, userspace).
The FRER standard does not mention any preferred location for the implementation.

A FRER implementation can benefit from the well-designed internal packet processing pipeline inside a hardware, especially in a TSN switch chip.
The maximum processing cycles for a packet can be calculated, with the extra cycles introduced by the FRER functions as well.
Therefore, the delay introduced can be bounded and deterministic, regardless of the packet rate or number of streams.

In the case of software implementation, FRER is just one application running on the system, competing for CPU resources with other applications.
Proper configuration of the system should be performed by the administrator, to ensure the FRER process can consume the required resources.
However, profiling and configuration can be difficult, in the kernel due to the large number of tunable parameters.
Implementing FRER in the Linux kernel, for example, requires careful design since Linux already has a very complex and performant Layer 2 stack.
FRER implementation proposals have been already made \cite{nxp_frer, frer_hanic}, but none of them accepted into the mainline Linux yet. 

Priority management and performance profiling are easier in userspace, however, the throughput decreased because of the copy between the kernel and userspace memory areas.
Usually TSN applications' transmitting bitrate is fixed and moderate, so throughput is less critical than determinism.
One can implement FRER with the help of the Linux userspace API-s, without caring about the design and operation of the Linux kernel's Layer 2 stack.

\section{The XDP FRER}

XDP \cite{xdp} is the eBPF \cite{ebpf} programmable packet processing framework of the Linux kernel.
It is essentially a packet pool that can be used by the network device's driver accompanied by one or more eBPF programs acting on the received packets.
With XDP, the decision about the fate of the packet (e.g. \textit{drop}, \textit{redirect}, or \textit{pass} to the network stack) happens as close as possible to the hardware.
This section goes through the eBPF/XDP toolset required for the implementation of FRER in XDP.

\subsection{Management}
Despite the XDP FRER data plane runs in the kernel, it is possible to communicate with the userspace through eBPF maps.
These maps can be accessed and updated both from the XDP program and the userspace management program.
XDP FRER uses these features to pass statistics to the management plane and to get VLAN translation tables, interface indexes, and configuration parameters from the management plane.
In this paper, VLANs are used in XDP FRER to distinguish streams and to separate background traffic from streams.
It is the best of both worlds; high flexibility and extensibility while kernelspace performance.

An XDP program is reference counted and freed by the kernel if all of its users disappear (e.g., every interface where the elimination function is attached goes down).
Even though the interfaces coming back, they no longer run the XDP program.
To avoid that, XDP FRER uses the BPF link API which automatically creates a file system reference for every XDP program.
As a result, XDP programs run as before when the interface is up again.

\subsection{eBPF packet manipulation}
The capability to add or remove packet header fields is necessary for the sequence generation and elimination functions.
The XDP FRER sequence generation function extends the head with the size of the R-tag, copies the Ethernet and VLAN headers, then inserts the R-tag with the new sequence number.
The elimination function is similar; it reads the sequence number from the R-tag, then copies the Ethernet and VLAN headers back, and finally shrinks the head of the packet.
Note: sequence generation can be skipped if the packet already has an R-tag.
In this case, replication can be performed without modifying the packet with sequence generation.
The same applies to elimination: it can leave the R-tag on the packet if configured that way.

\begin{listing}
	\caption{\label{lst:rtag}Example helper function pushing R-tag after the VLAN tag to an XDP frame}
\begin{minted}{c}
#define ETH_SIZE 14
#define VLAN_SIZE 4
#define RTAG_SIZE 6

static inline int add_rtag(struct xdp_md *pkt, uint16_t seq)
{
  /* Make room for R-tag (RTAG_SIZE)*/
  if (bpf_xdp_adjust_head(pkt, 0 - RTAG_SIZE))
    return -1;

  /* Checking the new boundaries for the sake of the verifier */
  void *data = (void *)(long) pkt->data;
  void *data_end = (void *)(long) pkt->data_end;
  if(data + ETH_SIZE + VLAN_SIZE + RTAG_SIZE > data_end)
    return -1;

  /* Move Ethernet+VLAN headers to the front of the buffer */
  /* memmove(destionation_addr, source_addr, size) */
  __builtin_memmove(data, data + RTAG_SIZE, ETH_SIZE + VLAN_SIZE);
  struct vlan_hdr *vhdr = data + ETH_SIZE;
  struct rtaghdr *rtag = data + ETH_SIZE + VLAN_SIZE;

  /* Prepare the R-tag */
  __builtin_memset(rtag, 0, RTAG_SIZE);
  rtag->nexthdr = vhdr->h_vlan_encapsulated_proto;
  vhdr->h_vlan_encapsulated_proto = bpf_htons(0xf1c1);
  rtag->seq = bpf_htons(seq);

  return 0;
}
\end{minted}
\end{listing}

For example Listing~\ref{lst:rtag}. show a helper function for pushing an R-tag.
The \texttt{bpf\_xdp\_adjust\_head} eBPF helper function can grow or shrink the head pointer of the packet.
In line~8, we extend the headroom with 6 bytes just to make enough space for the R-tag.
Then the \texttt{pkt->data} will points to the start of the extended headroom (line~12).

In line~19 we move the Ethernet and VLAN headers to this new start of the headroom, then in line~24-27 we fill the VLAN and R-tag header fields with proper protocol IDs and sequence number.

\subsection{Replication}

\begin{listing}
	\caption{\label{lst:repl}Example XDP routine implements FRER replication}
\begin{minted}{c}
struct tx_ifaces {
  __uint(type, BPF_MAP_TYPE_DEVMAP_HASH);
  ...
};

struct {
  __uint(type, BPF_MAP_TYPE_HASH_OF_MAPS);
  __array(values, struct tx_ifaces);
  ...
} repl_tx_map SEC(".maps");

SEC("xdp")
int replicate(struct xdp_md *pkt)
{
  void *data = (void *)(long) pkt->data;
  void *data_end = (void *)(long) pkt->data_end;
  if (data + ETH_SIZE + VLAN_SIZE > data_end)
    return XDP_DROP;

  int vid = get_vlan_id(pkt);

  /* "Identify" the stream by VLAN ID 
   * and lookup it's sequence generator function*/
  struct seq_gen *gen = bpf_map_lookup_elem(&seqgens, &vid);
  if (!gen)
    return XDP_DROP;

  /* Generate a sequence number and push an R-tag */
  uint16_t seq = generate_seq(gen);
  if (add_rtag(pkt, seq) < 0)
    return XDP_DROP;

  /* Replicate the packet for multiple egress interfaces */
  struct tx_ifaces *tx = bpf_map_lookup_elem(&repl_tx_map, &vid);
  if (!tx)
    return XDP_DROP;

  int flags = BPF_F_BROADCAST | BPF_F_EXCLUDE_INGRESS;
  return bpf_redirect_map(tx, 0, flags);
}
\end{minted}
\end{listing}

The implementation of the XDP FRER replication routine is presented in a simplified form in Listing~\ref{lst:repl}. 
Initially, the stream is identified by reading the VLAN ID of the packet (line~20).
Next, the sequence generator object for the given VLAN ID is retrieved (line~24).
Upon generating the new sequence (line~29), the R-tag gets pushed into the packet (line~30, see Listing~\ref{lst:rtag}).
In practice, sequence generation and replication can be decoupled.
For example, the node closer to the talker generates a sequence number for the packet, and the rest of the configured nodes do not touch the R-tag if the packet already has one, but only do the replication.

The \texttt{bpf\_redirect\_helper} eBPF function can be used with the \texttt{BPF\_F\_BROADCAST} flag (in lines~38 and 39) to transmit the same packet on multiple interfaces.
The special eBPF map type \texttt{BPF\_MAP\_TYPE\_DEVMAP\_HASH} (line~2) stores the egress interfaces identified by their index.
A map type named \texttt{BPF\_MAP\_TYPE\_HASH\_OF\_MAPS} is utilized to relate these maps of egress interfaces to the streams (VLAN IDs).
The lookup occurs at line~34.

\subsection{Elimination}

\begin{listing}
	\caption{\label{lst:elim}Example XDP routine implements FRER elimination functionality}
\begin{minted}{c}
SEC("xdp")
int eliminate(struct xdp_md *pkt)
{
  void *data = (void *)(long) pkt->data;
  void *data_end = (void *)(long) pkt->data_end;
  if (data + ETH_SIZE + VLAN_SIZE + RTAG_SIZE > data_end)
    return XDP_DROP;

  int vid = get_vlan_id(pkt);

  /* Lookup the sequence recovery object for the stream
   * by VLAN ID*/
  struct seq_recovery *rec = bpf_map_lookup_elem(&rcvy_map, &vid);
  if (!rec)
    return XDP_DROP;

  /* Read the sequence number of the packet and remove
   * the R-tag*/
  int seq = rm_rtag(pkt);
  if (seq < 0)
    return XDP_DROP;

  /* Protect the recovery object from concurrent
   * access (we keep receiveing packets)*/
  bpf_spin_lock(&rec->lock);
  bool pass = recover(rec, seq);
  bpf_spin_unlock(&rec->lock);
  if (pass != true)
    /* ...drop the replica packet(s)... */
    return XDP_DROP;

  /* Transmit the only packet that survives elimination */
  int *tx_ifindex = bpf_map_lookup_elem(&tx_map, &vid);
  if (!tx_ifindex)
    return XDP_DROP;
  return bpf_redirect(*tx_ifindex, 0);
}
\end{minted}
\end{listing}


Listing~\ref{lst:elim} shows a simplified version of the elimination function of XDP FRER.
Initially, we search for the sequence recovery object using the VLAN ID of the received packet (line~13).
The object stores the state of the recovery algorithm, which is essential for determining whether the packet is new, previously received, or potentially out-of-window.
We remove the R-tag in line~19, but first we read its sequence number.
In practice, this should be conditional because we may have further elimination functions in the network for this stream.

The sequence number (line~19) and recovery object (line~13) determine the fate of the packet.
It is important to protect the recovery object from concurrent access by using \texttt{bpf\_spin\_lock} (lines~25,~27).
This is necessary because the XDP elimination routine could be anywhere in the execution when a new packet of a member stream is received.
The implementation details of the \texttt{recover} function (line~26) are omitted due to space constraints.
However, the FRER standard \cite{1cb} defines  multiple recovery algorithms, our implementation use the \textit{vector recovery algorithm}.
The algorithm will accept the first received instance of the same packet and drop all subsequent replicas (line~30).
Finally, we lookup the egress interface for the packet which is passed by algorithm.

Furthermore, the elimination function has a recovery timeout (see the standard~\cite{1cb} for details).
If no packet is received after that interval, the elimination function resets its state and accepts the next received packet like it does after initialization, i.e., any sequence number is accepted.
That can be done either with BPF timers or by calling BPF program from userspace with the \texttt{BPF\_PROG\_RUN} infrastructure~\cite{bpf_run}.
XDP FRER uses the BPF test program method, since it does not have to be fast or very precise, which is called in every 2 seconds, and checks a per-recovery instance timestamp storing the time of the last received packet.

\section{Evaluation}

Since there are no software FRER implementations publicly available, we compared XDP FRER to our in-house userspace FRER implementation (referred to as uFRER in the following).
The uFRER uses the \texttt{AF\_PACKET} raw socket for packet reception and sending.
The \texttt{epoll} Linux API~\cite{epoll} is used for efficient socket monitoring.
Further optimizations are possible, e.g., using busy polling on the receiving socket or using \texttt{AF\_XDP} socket instead of \texttt{AF\_PACKET}.
However, we decided to stick with \texttt{epoll} and \texttt{AF\_PACKET} for the sake of simplicity.

\subsection{Testbed}
Our test machine is equipped with an AMD EPYC 7402P 24-Core Processor and 128Gb RAM.
Ten CPU cores are isolated from the Linux housekeeping tasks and regular process scheduling to avoid system noise while running the measurements on those cores.
We also modified the interrupt affinity, in order to direct the packet processing into the isolated CPUs (with \texttt{irqbalance} daemon disabled, to use static configuration).

CPU frequency governor set to performance, to keep all cores on fixed 2.8GHz, and avoid noise caused by the dynamic frequency scaling.
We used the Ubuntu 23.04 GNU/Linux distribution operating system, with its unmodified Linux kernel (version 6.2.0-27).
Furthermore, \textit{libbpf 1.2} and \textit{libxdp 1.3} development libraries were used to implement XDP FRER.

We used Intel i225 NICs for each measurement presented below.
\textit{igc} \cite{igc}, the device driver for this NIC, has been available upstream since Linux 4.20.
It is still receiving bug fixes and improvements, so all of our measurements reflect the driver version found in the above kernel version.

To verify that there is no major driver issue altering the results, we repeated all measurements with Virtual Ethernet (\textit{veth})\cite{veth} network devices.
We see consistent results except for slightly smaller minimum and average delays (since the packets don't actually leave the machine).
XDP has a native mode where the eBPF program is executed by the device driver.
We chose this mode for our measurements.
XDP also has a generic mode, which is executed by the network stack when the driver finished the reception of the packet, and allocates the socket buffer for it.
Generic Receive Offload (GRO) should be enabled on the \textit{veth} devices to utilize XDP native mode; otherwise, the XDP falls back to generic mode.

\begin{figure}[tp]
	\centering
	\includegraphics[width=0.85\columnwidth]{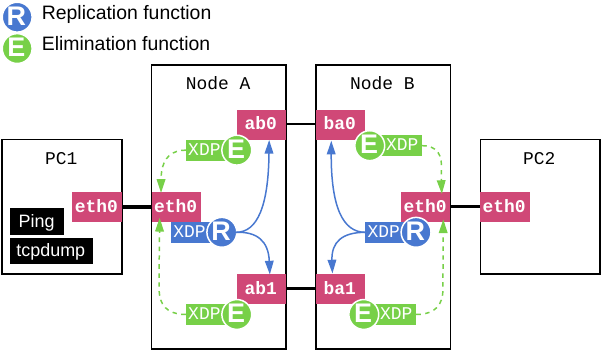}
	\caption{The test network used for experiments. The dashed lines from elimination functions mark the conditional packet drop}
	\label{fig:testbed}
\end{figure}

\subsection{Test network}
We built the simplest possible network for our tests: two hosts, \textbf{PC1} and \textbf{PC2}, each connected to a node (A and B) see Figure \ref{fig:testbed}.
There are two links used by FRER between the nodes.
The replication functions are placed at the \textbf{eth0} interfaces of nodes \textbf{A} and \textbf{B}, and the elimination functions are placed at the \textbf{ab0}, \textbf{ab1}, \textbf{ba0}, \textbf{ba1} interfaces of nodes \textbf{A} and \textbf{B}, respectively.
The actual hardware NICs are named differently by Linux, but we will stick with these more meaningful names for simplicity.

\subsection{Scenarios}
We tested the stream-protection capability and the performance of XDP FRER.
In the first case, we created artificial link failures to see if there was any packet loss.

We configured two streams for the performance test, one in each direction.
We used \textit{ping} (ICMP traffic) to test the RTT delay between the hosts.
The ping interval was set to 1 millisecond (1000 packets per-sec) and the packet size was 1000 bytes.
We generated 10000 packets with ping in each scenario.
We verified the values reported by ping for a handful of scenarios using the \textit{iperf2} UDP delay measurement method.
This verification confirmed that the accuracy of the ping is in the sub-ten microsecond range.

\subsubsection{Testing XDP FRER stream-protection from link failure}

\begin{figure}[tp]
	\centering
	\includegraphics[width=0.95\columnwidth]{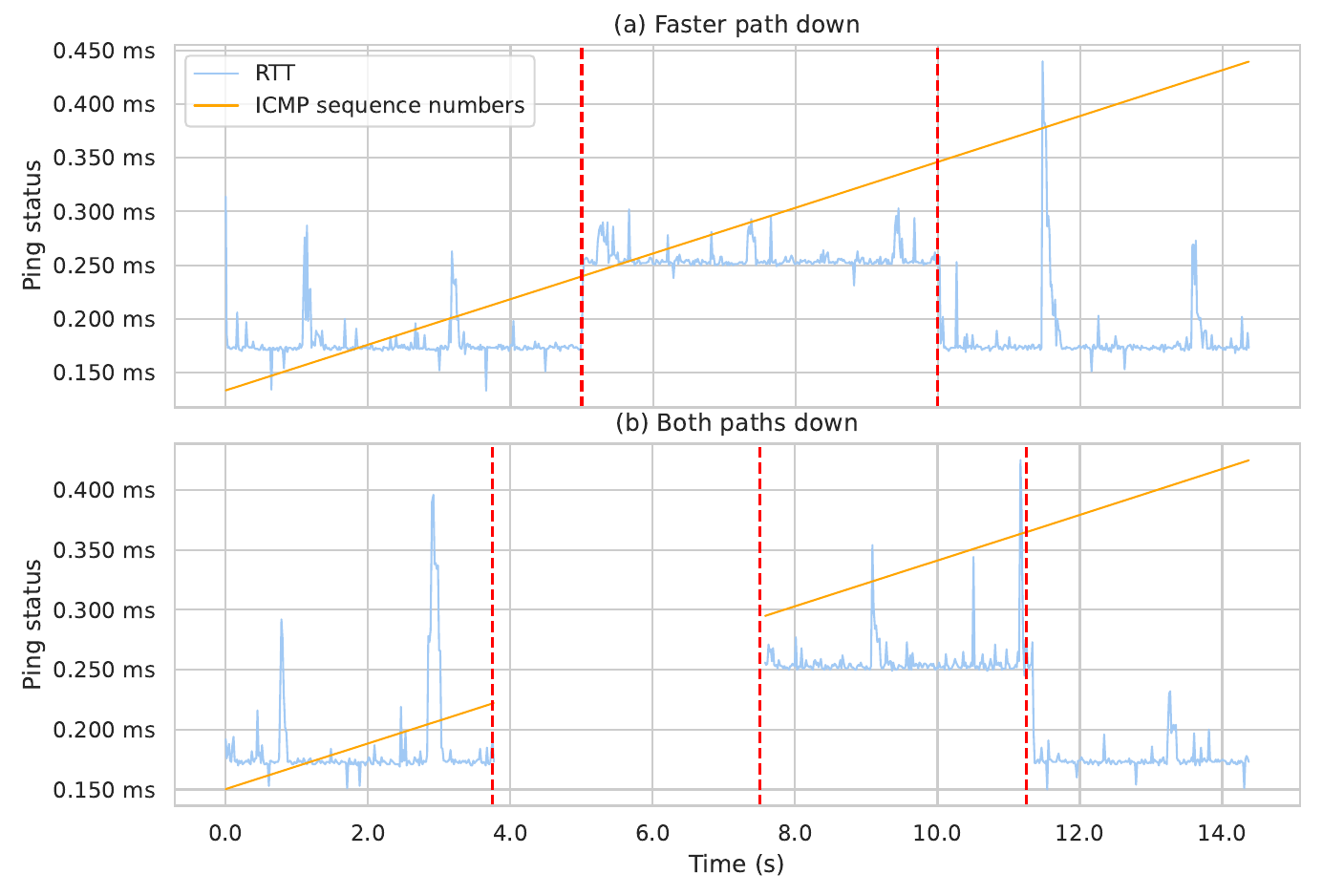}
	\caption{Ping RTT and sequence numbers with XDP FRER if one \textbf{(a)} or both \textbf{(b)} paths fails}
	\label{fig:error}
\end{figure}

In this scenario, we pinged PC2 from PC1 every 10ms.
After 5 seconds, we turned off \textbf{ab0-ba0} link for 5 seconds then turned it back on.
As visible on Figure \ref{fig:error}.a the ICMP sequence numbers on PC1 keep increasing since one path is still active.
To increase the visibility of the failure, we configured the i225 NICs to operate at 100Mbps on \textbf{ab0-ba0} path and at 2.5Gbps on the \textbf{ab1-ba1} path.
As a result, the failure becomes apparent in the RTT measurement because the failing path has a faster speed (with less frame serialization delay) and the alternate path has slightly higher delay.

Next, the test was repeated with both paths turned off around 4 sec (Figure \ref{fig:error}.b).
As expected, a gap is observed in the ICMP sequence numbers.
Then around 8 sec we turned path \textbf{ab0-ba0} and around 11 sec path \textbf{ab1-ba1} back on.
The ICMP sequence numbers started to increase immediately after the first path was turned on.

\subsubsection{Normal and Real-time priority}
We first investigated what is the effect of CPU load on the packet forwarding performance of uFRER.
We pinned the ping processes (on the PCs) and the two uFRER processes (nodes A and B) to different isolated CPU cores with the \textit{taskset} utility.
We generated an additional 100\% CPU load on the cores of the uFRER processes.
For that, we used the \textit{stress-ng}~\cite{stressng} utility with \textit{stream, matrix-3d} and \textit{cpu (loop)} stressors.

\begin{figure}[tp]
	\centering
	\includegraphics[width=0.95\columnwidth]{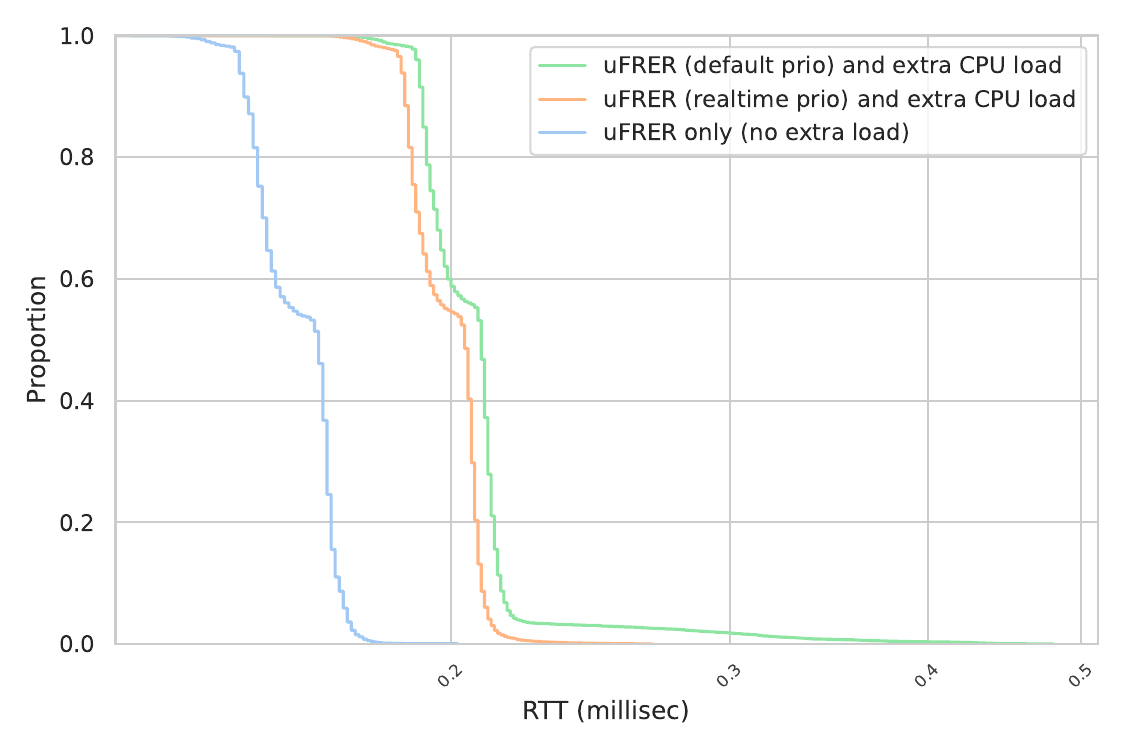}
	\caption{RTT distribution when running uFRER on isolated CPU core without extra CPU load (blue) and with extra load using normal (green) and real-time (orange) priorities}
	\label{fig:r2rt}
\end{figure}

The RTT distribution results are plotted in Figure~\ref{fig:r2rt}., without extra load (blue line) and extra CPU load (green line) on the uFRER's CPU core.
The measurements were repeated without additional load as well.
The load affects the forwarding delay of uFRER; the average latency increased slightly while the tail latency increased dramatically.

The simplest possible enhancement without any modification of the uFRER is running it with real-time priority.
Linux has had real-time priority support for more than a decade, designed especially for low-latency applications, e.g., audio-video equipment, and industrial controllers.
Important to note here that we used the default Linux kernel shipped with the distribution (Ubuntu), not a fully preemptable real-time kernel.
Running uFRER with the \textit{chrt} utility (\texttt{SCHED\_RR} scheduler with 99 task priority) decrease the average and tail forwarding latency as shown in Figure~\ref{fig:r2rt} with orange line.
In the following, we use real-time priority for uFRER in every scenario when comparing it to XDP FRER.

\subsubsection{XDP FRER and uFRER}
\begin{figure}[tp]
	\centering
	\includegraphics[width=0.75\columnwidth]{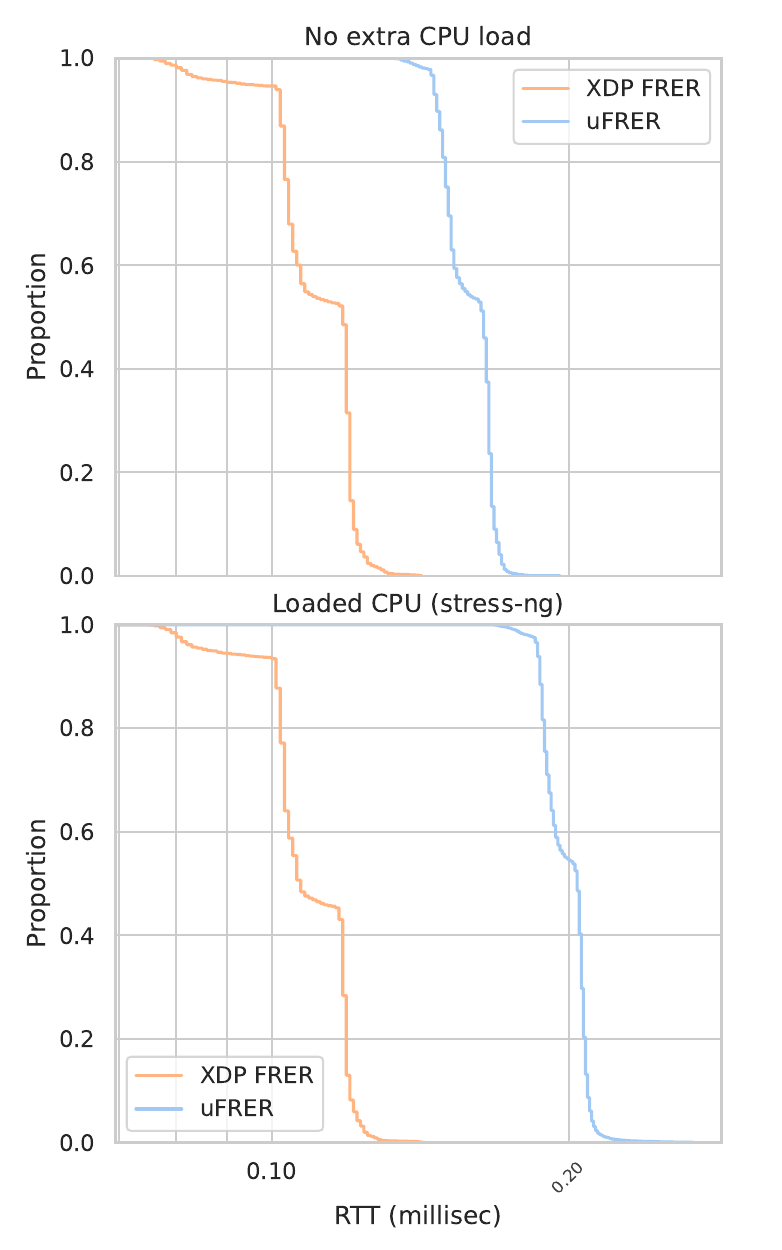}
	\caption{RTT distributions of XDP FRER and uFRER with and without extra CPU load}
	\label{fig:idlexdp}
\end{figure}

We compare the RTT distribution of XDP FRER and uFRER on idle and loaded systems in Figure~\ref{fig:idlexdp}.
As shown in the figure, XDP FRER does not only perform better in average and tail latencies, but it seems to be unaffected by the CPU load.
This can be explained by the traffic pattern.
1000 packets per-sec is a fairly low packet rate, at which almost every packet can be received in an interrupt context (IRQ).

The interrupt and software interrupt (called \textit{IRQ} and \textit{softIRQ}) is a high-priority, atomic, and non-preemptible context, or at least regular threads and softIRQs are always preempted by IRQs.
For this reason, CPU load does not affect XDP FRER as much as uFRER.
Important to mention is that since Linux 5.12 threaded NAPI is available.
In this mode, the NAPI (which is a poller abstraction for the network drivers to receive packets) runs as a kernel thread, and its resource allocation is controlled by the task scheduler.
This way, the CPU load can affect XDP FRER latencies more if the NAPI threads are not running at a high enough priority.

\subsubsection{XDP FRER under load}

We then added UDP background traffic generated by \textit{iperf} set to 1Gbps (10 UDP senders, 100Mbps each).
This extra traffic generates additional softIRQ load on the system.
XDP processing also takes place in the softIRQ context (unless threaded NAPI is explicitly enabled, more on that later).
If the softIRQ load is moderate, the CPU can keep up with the processing and also have time for other tasks scheduled for it.
However, in our case, the softIRQ load is high, and at some point the kernel needs to pause the softIRQ processing and postpone it until later.
Linux spawns a thread named \textit{ksoftirqd} (plus the CPU core ID) bound to each CPU core where the deferred softIRQs are offloaded.

\begin{figure}[t!]
	\centering
	\includegraphics[width=0.85\columnwidth]{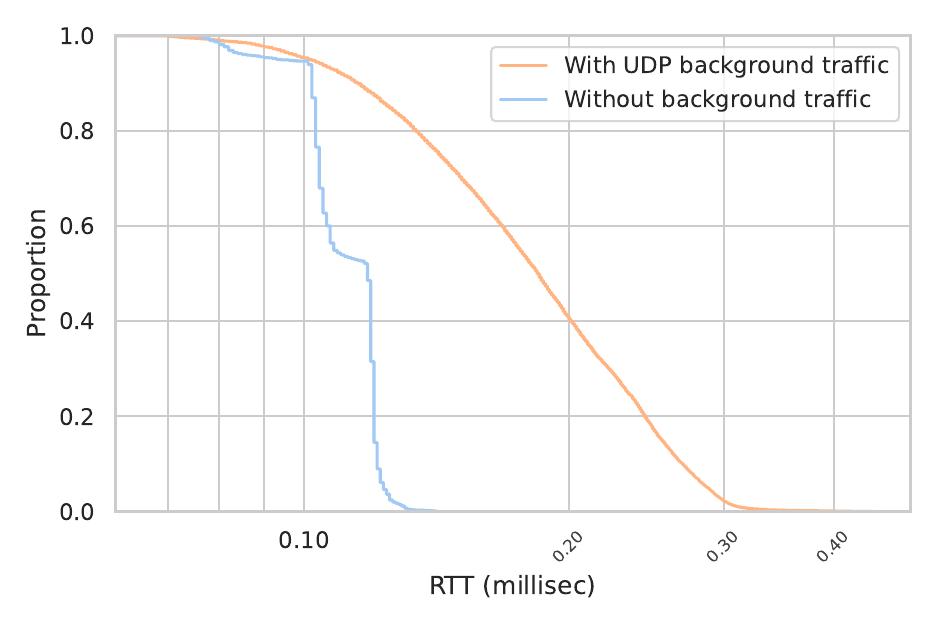}
	\caption{RTT distribution of ping (1000 PPS) with XDP FRER with and without UDP background traffic}
	\label{fig:loadedxdp}
\end{figure}

As we confirmed with \textit{bpftrace}, under moderate load, every packet is processed in the softIRQ context.
With background traffic, the CPU cannot keep up and offloads processing to \textit{ksoftirqd}.
While only about 1\% of the packets are processed in \textit{ksoftirqd}, the softIRQs still form a queue and cannot be processed immediately.
As can be seen in Figure \ref{fig:loadedxdp}. with background UDP traffic, the average and tail latency increased (orange line).

\section{Special note}



As mentioned earlier, a mutual exclusion (with \texttt{bpf\_spinlock}) required for accessing the recovery algorithm’s state of a given TSN stream, since the XDP FRER’s recovery function is triggered by packet arrivals.
This issue can be avoided by redirecting interrupts from multiple receiving NICs to the same CPU, since softIRQs are executed sequentially.
However, softIRQs run in a regular thread context with the real-time kernel, which makes them preemptible.
The same applies to threaded NAPI, which is available since Linux 5.12 (but disabled by default).
Using \texttt{bpf\_spinlock} does not limit the functionality of XDP FRER in any way.
However, it is currently not possible to call functions (even eBPF helper functions) between spinlocks.
In the future, eBPF verifier improvements will likely solve this limitation.

We replicated the physical test environment by using network namespaces and \textit{veth} interfaces during the prototyping of XDP FRER.
Although the topology and connections were essentially identical in both setup, it must be noted that \textit{veth} is purely a software network device, and the frames cannot be transmitted outside the machine.
After successful execution in the virtual environment, we ran the scripts on the physical setup.
It may be counterintuitive when attaching and running XDP program on a network interface, promiscuous mode is not implicitly enabled.
The packets are only sent to the CPU (and processed by XDP) if the NIC is in promiscuous mode, which can be enabled with the \textit{ethtool} command. On the other hand, in \textit{veth} environment promiscuous mode is the default.

\section{Conclusion}

This paper briefly explains the main concepts of per-packet service protection; in particular FRER, which is the standard Layer 2 variant.
We implemented FRER in eBPF/XDP and also in userspace. 
We have performed experimental analysis to compare the different implementations.
Our finding is that eBPF/XDP is a great fit for realizing FRER since it is as flexible as a userspace implementation considering management, whereas it can run in kernel space, hence can provide better performance.
Furthermore, no kernel modifications are required because it uses the standard eBPF interfaces and XDP packet processing hook points.
We also evaluated the performance of XDP FRER and showed that the introduced delay and low delay variation meet the requirements of deterministic communications.
In addition, we compared the performance of XDP FRER with that of our userspace FRER implementation and showed the advantages of XDP FRER on a loaded system with very low delay operation.

\bibliographystyle{ieeetr}
\bibliography{reference}

\end{document}